# Task-specific regularization loss towards model calibration for reliable lung cancer detection

Mehar Prateek Kalra¶†, Mansi Singhal†, and Rohan Raju Dhanakashirur§

*Dayalbagh Educational Institute (Deemed University), Agra, India*
*Dayalbagh Educational Institute (Deemed University), Agra, India*
*Indian Institute of Technology Delhi, New Delhi, India*

*Lung cancer is one of the significant causes of cancer-related deaths globally. Early detection and treatment improve the chances of survival. Traditionally CT scans have been used to extract the most significant lung infection information and diagnose cancer. This process is carried out manually by an expert radiologist. The imbalance in the radiologists-to-population ratio in a country like India implies significant work pressure on them and thus raises the need to automate a few of their responsibilities. The tendency of modern-day Deep Neural networks to make overconfident mistakes limit their usage to detect cancer. In this paper, we propose a new task-specific loss function to calibrate the neural network to reduce the risk of overconfident mistakes. We use the state-of-the-art Multi-class Difference in Confidence and Accuracy (MDCA) loss in conjunction with the proposed task-specific loss function to achieve the same. We also integrate post-hoc calibration by performing temperature scaling on top of the train-time calibrated model. We demonstrate 5.98% improvement in the Expected Calibration Error (ECE) and a 17.9% improvement in Maximum Calibration Error (MCE) as compared to the best-performing SOTA algorithm.*

## I. INTRODUCTION

[1]Lung Cancer is one of the major causes of cancer-related death worldwide. The mortality rate is higher than that of breast and prostate cancer combined [1]. This can be attributed to the lack of diagnosis and treatment in the early stages of cancer. Lung cancers can be categorized broadly into Squamous cell carcinoma, Adenocarcinoma, and Large cell carcinoma. Various diagnosing techniques have been developed to detect lung cancer. Computed Tomography (CT) is the most prominent technique used to diagnose infectious lungs and thereby detect cancer [2]. This process is generally carried out manually and therefore limits the scalability of diagnosis, resulting in lung cancer-related mortalities.

Many researchers have tried to automate the process of cancer detection. Narain Ponraj et.al [3] came up with a Local Optimal Oriented Pattern (LOOP) based CT image feature extraction method to automate the process of cancer detection. Though this was very popular, it required constant hyperparameter tuning and thereby reduced the scalability of the software. Jony et.al [4] tried resolving this issue by proposing a machine learning solution to it. They used a Marker-Controlled Watershed Gabor filter for segmentation of the CT images and later performed feature extraction on the same by using GCLM. The trained features were later learned using SVM. This technique reduced the dependency on hyperparameter tuning. Similarly, Wang et.al [5] used Random Forest to learn the features. However, these algorithms were not able to achieve the accuracy required.

Artificial Neural Networks (ANNs) were used by Kaur et.al [6]. They extracted GLCM [7] based features and used ANN to learn them. This was more of a heuristic way to use ANNs in solving the problem. Moradi et.al [8] came up with a 3D CNN-based model to solve the problem. Kanavati et.al [9] and chaunzwa et.al [10] use other standard CNN architectures, such as ResNet [11] and VGG [12], respectively, to learn cancer from CT images. Though these models worked considerably well in cancer detection, they were not so well accepted by the community due to model calibration issues.

Hence, it is very clear that the tendency of deep neural networks to commit overconfident mistakes makes them unusable for micro - suturing evaluation. In this paper, we propose a new technique to calibrate the model and restrict it from committing overconfident mistakes. Towards this,

---

[1]¶ *Corresponding Author:* Tel: +918889380023

† Bachelor Student with the Department of Electrical Engineering

§ Ph,d Student with Department of Computer Science Engineering

*E-mail addresses:* mehar.p.kalra2gmail.com, mansisinghal502@gmail.com, rohanrd28296@gmail.com

we claim the following contributions.

➔ We propose a new task-specific loss function to calibrate the neural network to reduce the risk of overconfident mistakes.

➔ We use the state-of-the-art Multi-class Difference in Confidence and Accuracy (MDCA) loss [13] in conjunction with the proposed task-specific loss function to achieve the same.

➔ We also integrate post-hoc calibration by performing temperature scaling on top of the train-time calibrated model.

➔ We demonstrate a 5.98% improvement in the Expected Calibration Error (ECE) and 17.9% in Maximum Calibration Error (MCE) as compared to the best performing generic loss function based SOTA algorithms

## II.  RELATED WORK

Model calibration is the most efficient way of solving this problem of overconfident mistakes [13]. The preliminary focus of model calibration is to convert the class probabilities into confidence. Therefore, in a calibrated model, the class probability of 70% indicates that the event is realized 70% of the time. Researchers have proposed various techniques to achieve model calibration. Post-Hoc Calibration and Train-Time Calibration are the two techniques that are mainly classified for calibrating DNNs. Post-hoc calibration comes up with the concept of a hold-out set to calibrate the model. On the other hand, the train-time calibration techniques modify certain system properties of the model to achieve calibration while training itself.

**Post-Hoc Calibration**

Post-hoc calibration utilizes the hold-out training set (validation set) to calibrate the model externally. Temperature scaling [14] uses a single scalar parameter T (>1) to rescale logit scores before applying the softmax function. Researchers use Temperature scaling as a variant of Platt scaling [15] to solve the same issue. Wenger et.al [16] use gaussian processes to model the latent space of the calibration function and thereby propose a non-parametric post-hoc calibration. Kuleshov et.al [17] propose to simplify the structure for the two-class classification problem. Xingchen et.al [18] propose a constrained optimization technique for well-controlled post-hoc calibration. They use mis coverage rate and convergence

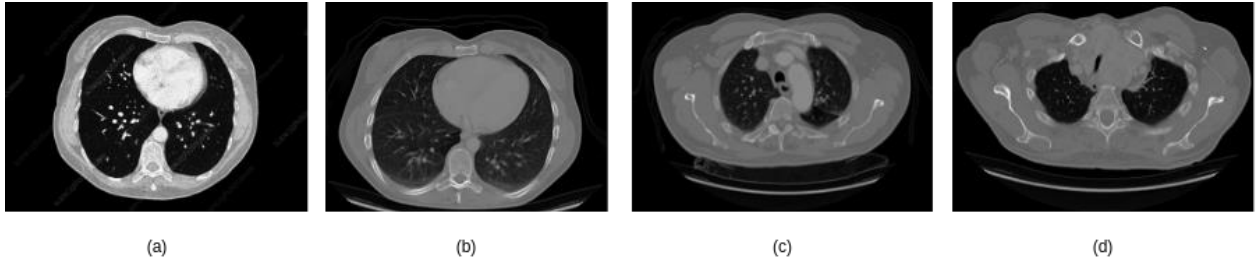

Figure 1: Sample images of Chest CT from each of the classes from the CCTSC dataset. (a) normal CT, (b) Adenocarcinoma, (c) Squamous carcinoma, (d) Large cell carcinoma

accuracy to obtain the constraints for the optimization. Each of these techniques calibrates the model in its own way. However, as the process of calibration is done after the training is complete, there is very little chance for the transformation of the outputs from class probabilities to confidence. This issue can be resolved by calibrating the model at the train time itself.

**Train-Time Calibration**

Brier et. al [19] introduces the Brier score to calibrate the binary probabilistic forecast. However, this system generally overfits due to its hard-bound calibration function and makes it unusable for real-world applications. Other approaches, such as Label Smoothing on soft targets [20] and entropy as regularization [21], are proposed to aid in improving calibration. Researchers have also considered using Focal loss [22] to calibrate the model by reducing KL-divergence and increasing the entropy simultaneously.

## III.  TASK SPECIFIC LOSS FUNCTION

In this paper, we solve the problem of lung cancer detection. We aim to obtain a model such that it outputs the confidence equivalent to the empirical frequency of its correctness. We use the publicly available Chest CT scan cancer (CCTSC) dataset to evaluate the proposed solution. The dataset contains CT scan images corresponding to four classes, namely: normal, Adenocarcinoma, Large cell carcinoma, and Squamous cell carcinoma. The dataset contains 928 CT scan images in total. Thus, the present dataset can be described as a set of samples $D = [(x_i, y_i)]_{i=1}^{928}$ coming from a joint distribution $D(X, Y)$ such that each of the samples $x_i \in X$ are the Chest CT images and $y_i \in Y = [0,1,2,3]$ denoting the ground truth labels. Formally, our problem can be defined as follows: Let there be a deep learning model predicting $y_p$ as the class label with $p_i$ as top-1 probability for the $i^{th}$ sample. The model is said to be calibrated if it obeys Equation 1.

$$L_{TS}(y_p, y_i) = \frac{1}{N} \sum_{i=1}^{N} (y_{i-norm} - y_{p-norm})^2 \quad (2)$$

where $y_{i-norm}$ for a $k$ class classification problem is as shown in Equation 3

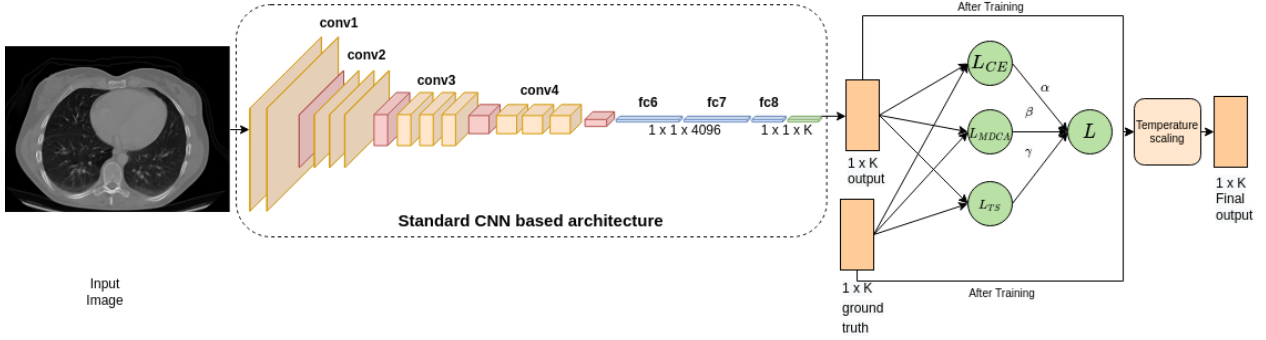

Figure 2: Proposed system for calibrated model towards lung cancer detection

$$P(y_p = y_i \mid p_i = s) = s \quad (1)$$

We propose to device a task-specific loss function $L_{TS}(y_p, y_i)$ that estimates the values of $p_i$ and $y_i$ which satisfies Equation 1.

It can be observed that our dataset does not span natural images. It is extremely restricted to the domain of chest CT. Since all the images show structural similarity to a great extent, the model must learn the differences in the finer parts of the image rather than learning the overall structure. Also, from the clinical expertise, it is evident that the normal CT scan comprises the entire lung, whereas each of the carcinomatous lung CT scans have some part of the lung covered by the tumor. In specific, squamous carcinoma has the least part of the lung covered by the tumor, followed by Adenocarcinoma and Large cell carcinoma.

Thus, if the ground truth labels are modified such that they obey the pattern of tumor deposition on the lungs (i.e-> Class 0 must correspond to normal Images, class 1 must be squamous carcinoma, class 2 must be Adenocarcinoma, and class 3 must be Large cell carcinoma), we can observe a gradation in tumor deposition as we progress from class 0 to class 3. Figure 1 encapsulates the visualization of this information.

This information of gradation can be exploited in training. i.e. the model predicting class 2 for a class 3 image can be penalized less as compared to a model predicting class 1 for a class 3 image. We encapsulate this information in the proposed loss function. The proposed loss function is shown in Equation 2.

$$y_{i-norm} = \frac{argmax\,(y_i)}{K-1} \quad (3)$$

A similar equation can be written for $y_{p-norm}$, indicating the completeness for Equation 2.

We add this loss function to the standard cross entropy loss function ($L_{CE}$) and the MDCA loss function ($L_{MDCA}$) [13] to achieve bin independent and class independent regularization. Therefore, the total loss function $L_{total}$ is given by Equation 4.

$$L_{total} = \alpha L_{CE} + \beta L_{MDCA} + \gamma L_{TS} \quad (4)$$

Where $\alpha, \beta, \gamma$ are constants, summing to 1.

We claim that the use of this loss function along with post-hoc calibration using temperature scaling gives the

calibrated model. Figure 2 shows the schematic diagram of the proposed system

### IV. IMPLEMENTATION DETAILS

We trained our model for 50 epochs on the CCTSC dataset and fine-tuned it on the same training dataset for another 50 epochs. The fine-tuning was divided into two steps, namely warm-up and final training. A batch size of 128 was chosen due to the high-performance ability of our GPU to load multiple high-resolution, multi-resolution images in a batch. For training on CCTSC, a learning rate

$1 \times 10^{-5}$, Adam optimizer with the momentum of 0.8, weight decay of $1 \times 10^{-4}$, and a gamma value of 0.1 was used. We fine-tuned using a learning rate of $1 \times 10^{-6}$, weight decay as 0.0005 and a momentum of 0.8. While fine-tuning, we only train the last 2 layers of the network, keeping the weights of the rest of the ResNet backbone frozen. All computations were carried out on a High Performance Computing Cluster having 80GB A100 GPUs. The training procedure took an average of 4 sec (approx.) per iteration
with a GPU memory occupancy of 31GB (approx.). The computations were carried out using Pytorch 1.11 and torchvision 0.12.0 libraries on Python 3.9 of Anaconda 3-2022.5.

Table 1 shows the performance of the proposed algorithm and the other SOTA algorithms on the dataset, using two different backbones, ResNet 34 and ResNet 50.

TABLE I
PERFORMANCE OF THE PROPOSED ALGORITHM AND THE OTHER SOTA ALGORITHMS

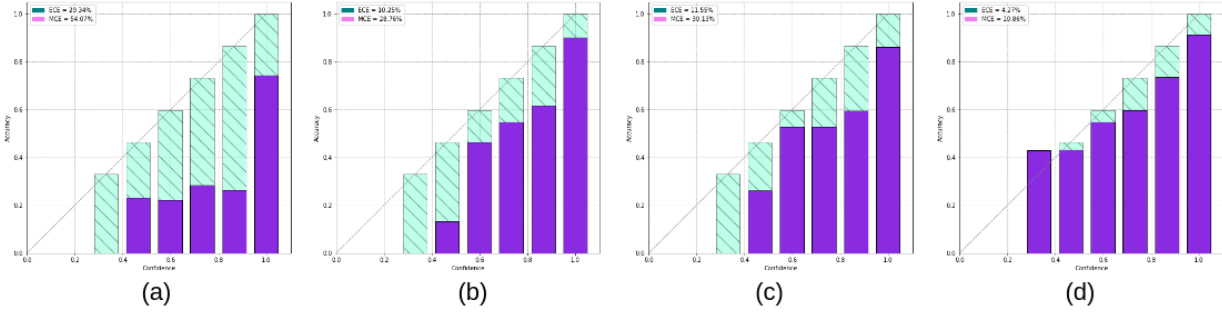

Figure 3: Comparison of the reliability plots for the proposed algorithm against SOTA implementations. (a) Cross Entropy [24] (b) Focal Loss [22] (c) MDCA Loss [13] (d) Proposed algorithm

## V. RESULTS AND DISCUSSION

We use 613 images of the dataset for training and 315 images for testing. Figure 4 shows the changes in the confidence and loss function over the iterations. It can be observed that the system converges before the maximum number of epochs is reached.

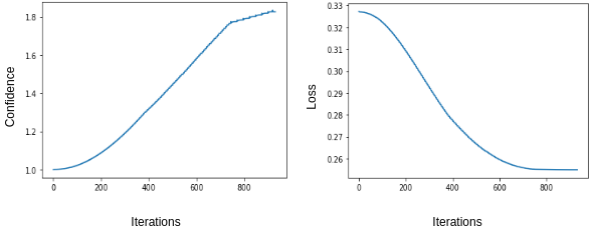

Figure 4: Change in confidence and loss value over the iterations.

We propose to evaluate the expected calibration score (ECE) and Maximum calibration score (MCE) for each of the models. MCE is the maximum difference between the accuracy and confidences averaged over each of the bins and is given by Equation 5.

$$MCE = max_{i \in \{1-B\}} |avg(p_i) - avg(s_i)| \quad (5)$$

ECE is the weighted average of the absolute difference between the confidence and accuracy of each bin. It shows the average error in calibration over the entire test set. Reduced values of ECE and MCE indicate better calibration.

| Architecture | ResNet 34 | | ResNet 50 | |
|---|---|---|---|---|
| | ECE | MCE | ECE | MCE |
| Cross Entropy [24] | 29.34 | 54.07 | 9.57 | 67.88 |
| Focal Loss [22] | 10.25 | 28.76 | 19.11 | 29.06 |
| MDCA Loss [13] | 11.55 | 30.13 | 21.76 | 30.71 |
| Proposed Algorithm | **4.27** | **10.86** | **4.65** | **6.69** |

It can be observed that the proposed algorithm performs better than the best performing SOTA by 5.98% in ECE and 17.90% in MCE for ResNet 34 backbone and 14.46% in ECE and 22.37% in MCE for ResNet 50 backbone.

TABLE II
ABLATION ANALYSIS OF THE PROPOSED ALGORITHM

| Architecture | ResNet 34 | | ResNet 50 | |
|---|---|---|---|---|
| | ECE | MCE | ECE | MCE |
| Only $L_{CE}$ | 29.34 | 54.07 | 9.57 | 67.88 |
| $L_{CE} + L_{MDCA}$ | 19.29 | 27.09 | 32.39 | 60.4 |
| $L_{CE} + L_{MDCA} + L_{TS}$ | 15.63 | 31.92 | 26.57 | 38.52 |
| Proposed Algorithm ($L_{CE} + L_{MDCA} + L_{TS}+$ | **4.27** | **10.86** | **4.65** | **6.69** |

| Temp scaling) | | | | |

This drastic improvement in the performance can be attributed to the fact that the other SOTA loss functions are designed for generic applications, whereas the proposed loss function is designed for this specific task. Although this comparison may not sound very fair, the unavailability of Lung cancer detection-specific loss functions forces us

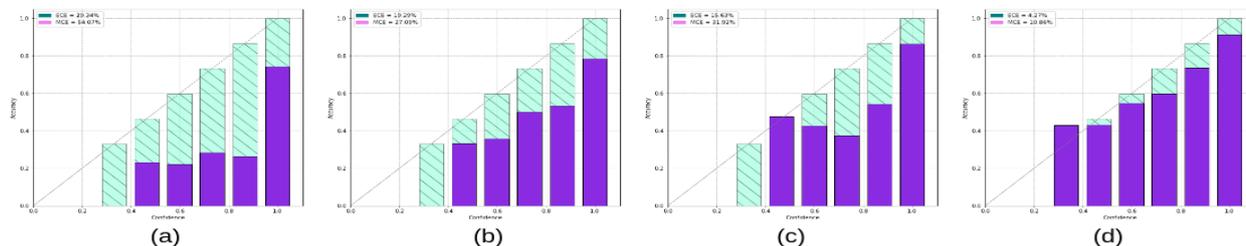

Figure 5: Reliability plot for ablation study. (a) is only cross entropy loss (b) is cross entropy with MDCA loss (c) is cross entropy with MDCA and TS loss (d) is the proposed algorithm

to have a comparison with the generic algorithms. The corresponding reliability diagrams for the ResNet 34 backbone can be seen in Figure 3.

We performed an ablation analysis by removing each of the components from the proposed system. The corresponding ECE and MCE values can be seen in Table 2, and the reliability plots for the ResNet 34 backbone can be seen in Figure 5. Table 2 confirms the need for each of the components. We also observe temperature scaling to be a significant component in model calibration.

## VI. CONCLUSION

In this paper, we proposed a new task-specific loss function to calibrate the neural network to reduce the risk of overconfident mistakes. We used the state-of-the-art Multi-class Difference in Confidence and Accuracy (MDCA) loss in conjunction with the proposed task-specific loss function to achieve the same. We also integrated post-hoc calibration by performing temperature scaling on top of the train-time calibrated model. We demonstrated a 5.98% improvement in the Expected Calibration Error (ECE) and a 17.90% improvement in Maximum Calibration Error (MCE) as compared to the best-performing SOTA algorithm. We also observed that the calibrated neural network is generally more reliable than the non-calibrated counterpart, and hence it can be used in critical applications such as lung cancer detection

## VII. ACKNOWLEDGMENT

We would like to express our sincere gratitude to our mentor and advisor, **Prof. Prem Kumar Kalra**. His constant support and guidance have made it possible for us to write this paper. Thank you, sir!